\documentclass[twocolumn]{aastex631}

\usepackage{CJK}

\graphicspath{{./}{figures/}}

\received{XXX}
\revised{YYY}
\accepted{ZZZ}

\usepackage{graphicx}	
\usepackage{amsmath}	
\usepackage{amssymb}	
\usepackage[inline]{enumitem}
\usepackage{gensymb}
\usepackage{array}
\newcolumntype{P}[1]{>{\centering\arraybackslash}p{#1}}
\newcolumntype{M}[1]{>{\centering\arraybackslash}m{#1}}
\usepackage{multirow}
\usepackage{tabularx}

\usepackage{xcolor}

\usepackage{color}

\shorttitle{FFPs from WD Kicks}
\shortauthors{Stephan \& Stassun 2026}

\usepackage{aas_macros}

\begin{document}

\title{Contribution of White Dwarf Formation Kicks to the Free-Floating Planet Population}

\author[0000-0001-8220-0548]{Alexander P. Stephan}
\affiliation{Department of Physics and Astronomy, Vanderbilt University, Nashville, TN 37235, USA}

\author[0000-0002-3481-9052]{Keivan G. Stassun}
\affiliation{Department of Physics and Astronomy, Vanderbilt University, Nashville, TN 37235, USA}


\begin{abstract}

Free-Floating Planets (FFPs) are a distinct class of exoplanets that do not orbit stars but are nevertheless found to be very common. A variety of formation mechanisms have been proposed as their origin, such as "star-like" direct collapse from gas and dust clouds or ejection from young planetary systems via dynamical instabilities. Here, another possible formation scenario is explored that would instead apply for old planetary systems, in the form of White Dwarf (WD) formation kicks. Observations over recent years have shown that WDs experience a mild recoil kick during their formation from Asymptotic Giant Branch (AGB) stars. Here we show that, while WD formation kicks directly unbind only $\sim1\%$ of the known planet and exoplanet population, they drive dynamical instabilities in $\gtrsim40\%$ of known long-period multi-planet systems, likely leading to planet ejection and FFP generation by most WDs. We also show that FFPs generated via WD kicks will have undergone potentially significant and long-lasting heating via their host stars' enhanced AGB luminosities. Given the low ejection velocities due to the weakness of the WD kicks, such warmed FFPs can thus be associated with their former host stars for several Myr after formation. Therefore, WD formation kicks contribute a distinct, observationally identifiable FFP sub-population comprising a few percent of the Galaxy's FFPs, relevant for the design of future FFP surveys.

\end{abstract}



\section{Introduction} \label{Introduction}

Over the last few decades, a number of observational surveys and campaigns, very prominently involving space-based instruments like {\it Kepler} or {\it TESS}, have discovered thousands of exoplanets, orbiting stars of all evolutionary phases and with a wide range of stellar masses, in a large variety of orbital architectures often very unlike our own solar system \citep[e.g.,][]{WolszczanFrail1992,Charpinet+2011,Howard+2012,Gettel+2012,Vanderburg+2020}. While observational techniques like transit detection or radial velocity measurements have explicitly looked for exoplanets bound to stars, some observational methods, such as microlensing or direct imaging, have been able to uncover a growing population of ``free-floating'' planets (FFPs) that are not bound to any specific star \citep[e.g.,][]{Sumi+2011,Gaudi+2012,Udalski+2015}. The soon-to-be-launched {\it Roman} Space Telescope \citep{Spergel+2015,Akeson+2019} is expected to find hundreds, if not thousands, of FFPs over the course of its Galactic Exoplanet Survey via microlensing \citep{Barclay+2017,Johnson+2020}, which will greatly improve the available population statistics of this type of exoplanet.

\begin{figure*}
    \centering
    \includegraphics[width=1\linewidth]{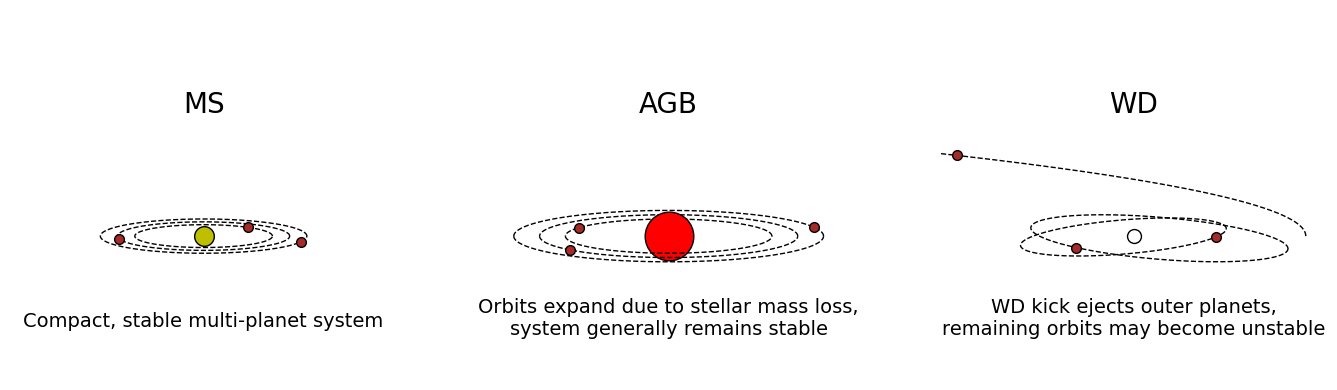}
    \caption{{\bf Schematic of the Evolution of a Planetary System due to Stellar Mass loss and WD Kicks.} 
    The figure depicts a multi-planet system, viewed nearly edge-on, that is initially compact, co-planar, and stable during its host star's MS phase (left frame). During the AGB phase, the orbits expand but generally remain stable (middle frame). After becoming a WD and experiencing a WD formation kick, outer planets may become unbound, while the remaining planetary orbits obtain new eccentricities and inclinations, potentially triggering instabilities (right frame).}
    \label{fig:summary_plot}
\end{figure*}

The theoretical origin of FFPs has been studied extensively using a variety of models, including as a byproduct of planet-planet scattering \citep[e.g.,][]{RasioFord1996,Chatterjee+2008,BN2012,VerasRaymond2012,BhaskarPerets2025}, in-situ formation \citep[e.g.,][]{Boss2001,Krumholz+2005}, or inter-cluster scattering and stellar flybys \citep[e.g.,][]{Spurzem+2009,Malmberg+2011,Parker+2012}, where much focus has usually been given to the formation phase of exoplanetary systems. However, it is suspected that various mechanisms, such as long-term dynamical instabilities or stellar evolution effects, can also unbind exoplanets from their home stars \citep[e.g.,][]{Veras+2011,Veras+12}. In this work the contribution to FFP formation by a particular stellar evolution effect, the recoil kicks associated with the formation of White Dwarf (WD) stars, is explored.

Observations of WDs in binary systems with {\it Gaia} have revealed that WDs undergo a relatively weak kick (on the order of $\sim0.75$~km~s$^{-1}$) during their formation \citep{ElBadryRix2018}, which is further supported by an observed dearth of WDs in stellar clusters \citep[e.g.,][]{Fellhauer2003,Heyl2007,Heyl2008a,Heyl2008b,Davis+2008,Fregeau+2009}. While the exact stellar mechanism of this kick is uncertain (i.e., whether the kick is the result of an instantaneous mass loss event, or of the cumulative effect of many small mass loss events, or of a more gradual asymmetric mass loss episode during its entire post-main sequence evolution), its impact on planetary and binary system dynamics has been shown to be significant \citep[e.g.,][]{Shariat+2023,Shariat+2024,Stephan+2024,Shariat+2025,OConnor2026}.

A common outcome of WD kicks will be the ejection of companions, both stellar and planetary, with a higher likelihood of ejection for wider-orbit companions, while the companions that remain bound will exhibit altered orbital eccentricities and inclinations that can potentially lead to instabilities and further ejections (see Fig~\ref{fig:summary_plot} for reference). In this work the fraction of exoplanets that can be expected to become FFPs following the observed characteristics of WD kicks as described by \citet{ElBadryRix2018} is determined. Additionally, it is investigated how well and for how long FFPs produced by WD kicks can be associated with their original WD hosts. The results provide important guidance for analyzing future FFP population statistics, and may inform the design of future direct imaging FFP surveys to complement surveys based on microlensing, such as the upcoming {\it Roman} Galactic Exoplanet Survey \citep{Penny+2019,Wilson+2023}.


\section{WD Kick Effects}

The WD formation kick empirically determined by \citet{ElBadryRix2018} has a velocity $v_{\rm kick}$ that follows a Maxwellian distribution with the probability function
\begin{equation}\label{eq:maxwell}
    P(v_{\rm kick}) = \sqrt{\frac{2}{\pi}}\frac{v_{\rm kick}^2}{\sigma_{\rm kick}^3}\exp{\left[-\frac{v_{\rm kick}^2}{2\sigma_{\rm kick}^2}\right]} \ ,
\end{equation} 
with $\sigma_{\rm kick}\approx0.5$~km~s$^{-1}$ and with the peak of the distribution at $v_{\rm kick}=\sqrt{2}\sigma_{\rm kick}=0.75$~km~s$^{-1}$. At the moment, the exact stellar mechanism that generates the kick is uncertain. As such, here it is assumed that the kick occurs instantaneously at the end of the host star's Asymptotic Giant Branch (AGB) phase, in an isotropically random direction. A caveat here is that some recent works suggest that the kick may instead be the cumulative result of a large number of small kicks during the AGB phase \citep{Fuller2026} or that the kick is indeed much stronger in certain circumstances \citep{Chawla+2025}, which may lead to different dynamical evolution pathways than which we focus on in this work.

\subsection{Orbit Separation due to WD Kicks}

To completely unbind a planet via a kick, the combination of the orbital velocity $v_{\rm orb}=\sqrt{GM/a}$ (with $G$ being the gravitational constant, $M$ the combined mass of star and planet, and $a$ the semi-major axis (SMA) of the planet, assuming a circular orbit) and the kick velocity $v_{\rm kick}$ must surpass the escape velocity $v_{\rm esc}= \sqrt{2}\ v_{\rm orb}$ such that 
\begin{equation}\label{eq:escape}
    v_{\rm esc}  < |\vec v_{\rm orb} - \vec v_{\rm kick}|.
\end{equation}
Note that the kick acts on the star, thus a kick in the direct opposite direction of the orbital velocity will increase the total orbital speed. As long as $(\sqrt{2}+1)v_{\rm orb}\geq v_{\rm kick}\geq(\sqrt{2}-1)v_{\rm orb}$, one can define the minimum angle $\theta_{min}$ between the orbital motion and the kick direction below which all angles $\theta$ will lead to orbital separation, such that \begin{equation}\label{eq:theta}
    -1 \leq\cos{\theta}\leq\cos{\theta_{min}} = \frac{v_{\rm kick}^2 - v_{\rm orb}^2}{2v_{\rm orb}v_{\rm kick}}\leq1.
\end{equation}Naturally, kick velocities above $(\sqrt{2}+1)v_{\rm orb}$ will lead to orbit separation for all angles $\theta$.
When assuming an isotropically random kick direction, the angle $\theta$ is distributed uniformly in cosine, thus allowing calculation of the kick separation fraction $F_{\rm sep}$ directly from Eq.~\ref{eq:theta}, such that
\begin{equation}\label{eq:frac}
    F_{\rm sep} = \frac{1+\cos{\theta_{min}}}{2}=\frac{1}{2}+\frac{v_{\rm kick}^2 - v_{\rm orb}^2}{4v_{\rm orb}v_{\rm kick}},
\end{equation}
limited to a minimum and maximum value of $0$ and $1$, respectively.

\begin{figure}
    \centering
    \includegraphics[width=\linewidth]{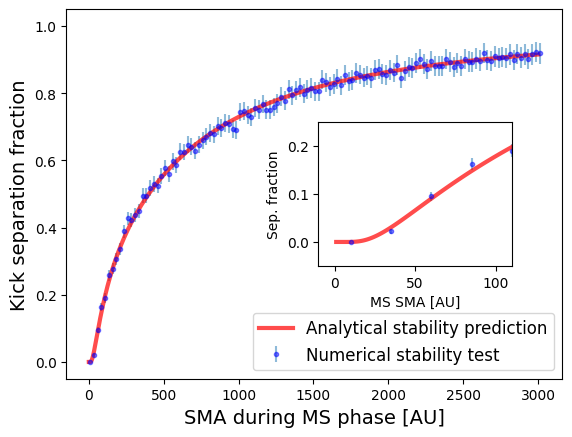}
    \caption{{\bf Fraction of Kick-separated Planetary Orbits vs. initial SMA.} The kick distribution of \citet{ElBadryRix2018} is applied to the separation fraction calculated in Eq.~\ref{eq:frac}, resulting in the average expected separation fraction for a given planetary orbital SMA (red curve; the blue dots are from a numerical test). Note that, to allow direct comparisons to planetary systems orbiting MS stars, the SMA's shown are from the MS lifetime of the host star, before significant mass loss has occurred, while the kick is applied at the end of the AGB phase, after most mass loss has already taken place and the SMAs have expanded accordingly. In this example, the host star had an initial mass of $1$~M$_{\odot}$, which reduced to $0.52$~M$_{\odot}$ by the end of the AGB phase. The inset focuses on orbits closer than $\sim100$~AU, relevant for the outer solar system planets.} 
    \label{fig:ffp_kick}
\end{figure}

Eqs.~\ref{eq:maxwell} and \ref{eq:frac} can be combined to calculate the expected fraction of planets that will undergo orbital separation for a given stellar mass and orbital semi-major axis. An example is shown in Fig.~\ref{fig:ffp_kick}, where the validity of the analytical kick separation equations outlined above is also tested against numerical calculations.

Fig.~\ref{fig:ffp_kick} highlights that the kick separation percentage approaches $100\%$ for wider orbits, surpassing $70\%$ by SMAs of about $1000$~AU, for a solar-type stellar host. For the solar system planets the separation fractions are generally small, only reaching about $2\%$ for Neptune-like orbits. However, examples of discovered exoplanets with wider orbits exist, such as in the case of the HR 8799 system, which has an outer confirmed planet, HR 8799 b, with an SMA of about $67$~AU \citep{Wang+2018}. HR 8799 is a young star at the F and A-type boundary, with a mass of $\sim1.4$~M$_{\odot}$, which is estimated to reduce to $\sim0.56$~M$_{\odot}$ by the time it becomes a WD, based on calculations performed with the stellar evolution code {\tt SSE} \citep{Hurley+00}. As such, HR 8799 b is estimated here to have a kick separation likelihood of about $15\%$.

\subsection{WD Kicks triggering Orbital Instabilities}\label{subsec:inst}

While the previous calculations showcase that WD kicks will unbind a non-negligible fraction of planets, the separation likelihood is small for most known planets and exoplanets. This is partially due to limitations of the current main exoplanet detection methods, namely transits and radial velocity, which generally favor short-period orbit detections \citep[e.g.,][]{Borucki+2011,Sabotta+2021}. However, kicks may lead to the loss of planets from a system not only via immediate unbinding, but also by triggering orbital instabilities and scattering events in multi-planet systems. 

For planets that remain bound, the WD kick will alter their SMAs, eccentricities, inclinations, and arguments of periapsis. The changes of orbital parameters due to kicks and their effects have been explored in various works \citep[e.g.,][]{Kalogera2000,LuNaoz2019,Shariat+2023,Stephan+2024}. As such, in a given multi-planet system, some fraction of neighboring pairs of planetary orbits may become unstable. To characterize whether a given system will become unstable beyond using a direct n-body simulation, one can use various indirect metrics: 

(1) If two planetary orbits start to cross (where the outer orbit's periapsis is closer to the star than the inner orbit's apoapsis), one generally expects there to eventually be a close encounter, unless the orbits are in a resonance (such as for Neptune and Pluto). 


(2) \citet{Petrovich2015b} numerically determined a stability criterion similar to one by \citet{Mardling+01}, valid for low mutual inclinations. It is defined as \begin{equation}
    \label{eq:petrovich}
    \frac{a_2(1-e_2)}{a_1(1+e_1)}>2.4\left[max\left(\frac{m_1}{M_*},\frac{m_2}{M_*}\right)\right]^{1/3}\sqrt{\frac{a_2}{a_1}}+1.15.
\end{equation} Systems that violate this criterion can be expected to eventually lead to planetary ejections or collisions with the central star, even if the orbits do not cross.

\begin{table}[htbp]
    \centering
    \caption{{\bf Likelihood Estimates of Instability after the WD Kick for Neighboring Pairs of Solar System and HR 8799 Planets.} The WD kick identified by \citet{ElBadryRix2018} is applied to the planets in these systems with random orientation of the kick vector and at random positions of the planets along their orbits. Neighboring pairs of orbits are examined to determine the likelihood that they violate the stability criteria outlined in Sec.~\ref{subsec:inst} post WD kick, namely (1) orbit crossing or (2) the criterion by \citet{Petrovich2015b}. Also shown is the percentage of cases where at least one member of the pair is directly ejected via the kick. The uncertainty for all listed values is about $1\%$.}
    \hspace*{-2.5cm}
    \begin{tabular}{|l||c|c|c|}
    \hline
      Planet Pair & (1) & (2) & Kick\\
      \hline
      \hline
       Venus-Earth & $4\%$ & $15\%$ & $0\%$ \\
      \hline
       Earth-Mars & $7\%$ & $22\%$ & $0\%$\\
      \hline
      Mars-Jupiter & $0\%$ & $0\%$ & $0\%$ \\
      \hline
      Jupiter-Saturn & $15\%$ & $34\%$ & $0\%$ \\
      \hline
      Saturn-Uranus & $21\%$ & $22\%$ & $0\%$ \\
      \hline
      Uranus-Neptune & $53\%$ & $50\%$ & $2\%$ \\
      \hline
      \hline
      HR 8799 e-d & $46\%$ & $82\%$ & $2\%$ \\
      \hline
       HR 8799 d-c & $54\%$ & $77\%$ & $5\%$ \\
      \hline
       HR 8799 c-b & $50\%$ & $64\%$ & $15\%$ \\
      \hline
    \end{tabular}
    \label{tab:instability}
\end{table}

\begin{figure}
    \centering
    \includegraphics[width=\linewidth]{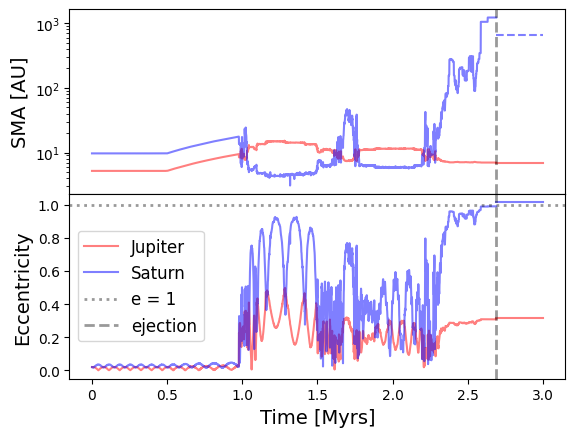}
    \caption{{\bf Example Evolution of a 2-Planet System undergoing a WD Kick.} Shown is the orbital evolution of the SMAs (upper frame) and eccentricities (lower frame) of a Jupiter-like (red lines) and Saturn-like (blue lines) planet orbiting a solar-type star, as modeled using {\tt REBOUND}. The orbits are initially nearly coplanar and circular, and are dynamically stable. The orbits remain stable throughout the RGB mass loss phase (starting at about $0.5$~Myr in the simulation; the mass loss phase is sped up compared to a realistic stellar evolution simulation to reduce computation time), but \textcolor{black}{a $1$~km~s$^{-1}$ strong} WD kick at the end of the RGB phase \textcolor{black}{(at about $1$~Myr)} majorly alters the planetary orbital SMAs and eccentricities, triggering instability. Over the course of the next $\sim2$~Myr, the planets undergo repeated scattering encounters, until the outer Saturn-like planet is ejected, becoming a FFP \textcolor{black}{(moment of ejection marked with gray dashed line). The SMA of the ejected planet is about $-670$~AU (shown by the blue dashed line, as an absolute value due to the logarithmic y-axis), implying an excess ejection velocity of about $0.85$~km~s$^{-1}$.}} 
    \label{fig:rebound_kick}
\end{figure}

As example cases, Tab.~\ref{tab:instability} shows the percentages of unstable post-kick orbital configurations, as estimated by the criteria outlined above, applied to the solar system and the HR 8799 system. A numerical test using the n-body code {\tt REBOUND} \citep{ReinLiu2012,ReinSpiegel2015} was also performed for some of the listed example configurations (see also Fig.~\ref{fig:rebound_kick}), confirming that systems with crossing orbits or that are violating the \citet{Petrovich2015b} criterion will in general lead to planetary ejection over timescales of a few $10^6$ years or less. As such, the instability likelihood estimated with the \citet{Petrovich2015b} criterion appears to be a reliable approximation.

As the examples in Tab.~\ref{tab:instability} highlight, orbital parameter changes induced by the WD kick can make a significant fraction of planetary systems unstable, leading to a much higher yield of ejected exoplanets as FFPs than from the WD kick unbinding exoplanets directly. A caveat, however, exists in the form of mutual inclinations, and instabilities triggered by the presence of more than $2$ planets or secular resonances, which the stability criteria do not include consistently, and which may significantly alter the actual stability of the systems. As such, the given percentages are conservative estimates that can be further refined using numerical n-body simulations.



\section{WD-Kick FFP Yield}\label{sec:yield}

In order to estimate the combined contribution of direct orbital separation via WD kicks and eventual planet ejection via kick-triggered orbital instabilities to the overall production of FFPs, one has to estimate the distribution of exoplanet system orbital parameters, ideally as a function of host star mass and planet mass. This is difficult due to the significant incompleteness in the exoplanet data in respect to small planets and wide orbits. As such, however, one can still use the currently known population of exoplanets to estimate a conservative lower limit for FFP production from WD kicks.

There are currently about $6150$ confirmed exoplanets listed on the {\tt NASA Exoplanet Archive} \citep{Akeson+2013} (accessed in March 2026), however only about $5803$ have reliable estimates for host star mass, exoplanet mass, and exoplanet orbit SMA. Of these, about $3341$ exoplanets do not have any known exoplanet companions, and $2462$ are found in multi-planet configurations of $2$ or more exoplanets. Additionally, only $2276$ of these exoplanets orbit MS stars with masses of $1$~M$_\odot$ or larger, which is approximately the smallest type of stars that have had time to evolve into WDs by the current age of the universe. Only the exoplanets that are orbiting these ``WD progenitors'' are considered for estimating system stability against WD kicks. As an additional restriction, planetary orbits that would be engulfed during the AGB are also discarded. Given the short-period biases in the known exoplanet population, these cuts eventually reduce the number of multi-exoplanet systems in consideration to just $53$. To reiterate, the small number of viable systems is mostly a reflection of the current observational biases in the known exoplanet catalog, and one should expect longer period multi-planet systems to be common.

For each system, {\tt SSE} is used to estimate the host star's eventual WD mass. Each exoplanet is tested for direct separation via the WD kick following Eqs.~\ref{eq:maxwell} and \ref{eq:frac}, using the orbital parameters as listed on the {\tt NASA Exoplanet Archive} (where no orbital eccentricity is known, the orbit is assumed to be circular). For exoplanets with companions, Eq.~\ref{eq:petrovich} is used to estimate instability, which is assumed to lead to FFP ejection. Using {\tt REBOUND}, the instability estimate is also checked against numerical n-body integration to account for the effect of mutual inclinations, limited to $5$~Myr in post-kick duration due to computational limitations. For both kick- and planet-planet-scattering induced planet ejection, we record the hyperbolic excess velocity $v_\infty=\sqrt{-GM/a}$ (where $a$ is negative for hyperbolic orbits), which gives the speed at which the ejected FFP and its now former host star will separate.

The WD kick can directly unbind only about $1.3\%$ of the known exoplanet population, which is in line with expectations given the observational bias towards short periods, many of which may even be engulfed during the AGB phase. However, multi-planet systems appear to become unstable post-kick in nearly $41\%$ of cases based on Eq.~\ref{eq:petrovich}. In the numerical test with {\tt REBOUND}, $25$ out of the $53$ simulated systems, or $\sim47\%$, lead to ejections within $5$~Myr of the kick (mostly within the first two Myr; see also Fig.~\ref{fig:rebound_kick}), broadly consistent with Eq.~\ref{eq:petrovich}. These results show that WD kicks can lead to the ejection of FFPs from a large fraction of exoplanet systems. In the most optimistic scenario, where most stars have multi-planet systems with wider orbits than currently observed, it can be assumed that most WDs that had exoplanet systems will have ejected one or more FFPs due to their WD formation kick or the following orbital instabilities.

\textcolor{black}{In our numerical experiments, we observe that the typical excess ejection velocity for FFPs directly unbound by the WD kick is $\sim1$~km~s$^{-1}$. For FFPs generated via planet-planet scattering, the excess ejection velocities are generally higher, with a most likely excess ejection velocity of $\sim1.5$~km~s$^{-1}$. In previous works studying FFP generation from planet-planet scattering during the host star's main sequence lifetime \citep{BhaskarPerets2025}, the excess ejection velocity appears to inversely correlate with the SMA of the innermost planet in a given system, and as the systems that we consider do not have very short-period planets remaining due to AGB-phase engulfments, one would expect average excess ejection velocities of $\sim2$~km~s$^{-1}$ or less \citep[see Fig.~6 in][]{BhaskarPerets2025}, consistent with our results. Given that we expect most WD-kick FFP generation to come from post-kick orbital instabilities and scattering, rather than direct orbital separation, we assume an overall expected excess ejection velocity of $\sim1.5$~km~s$^{-1}$ for the FFPs generated following a WD kick.}


To put the WD kick contribution estimate into perspective, the FFP population generated during the formation phase of planetary systems is estimated to be on the same order as the number of bound exoplanets \citep[e.g.,][]{Mroz+2017,Gould+2022,Sumi+2011}. Assuming about $4\times10^{11}$ stars and given an estimated number of about $10^{10}$ WDs in the Galaxy \citep[e.g.,][]{Napiwotzki2009}, one can thus estimate that WD-kick produced FFPs can contribute on the order of a few percent to the total FFP population.


\section{Observable WD-Kick FFP Characteristics}

While FFPs generated due to WD kicks do not dominate the overall FFP population, WD-kick FFPs will have several characteristics that can serve as distinguishing observational markers: \begin{enumerate}
    \item Given the modest expected ejection excess velocities of $\sim1.5$~km~s$^{-1}$ as discussed in Sec.~\ref{sec:yield}, ejected FFPs will remain close to their former home systems for some time. On average, the FFPs will travel about $1.5$~pc/Myr relative to their former host stars. 
    \item FFPs originating from WD kicks will have evolved for a long time in the close proximity of a star and usually a full planetary system. This may have resulted in more weathering compared to FFPs ejected from young systems, due to repeated asteroid impacts and interactions with the stellar wind, especially during the AGB phase, with various implications for the FFPs' surface conditions. The intense radiation during the AGB may also significantly increase the FFP temperature for an extended period of time.
    \item As the galaxy continues to age, more stars will evolve into WDs and undergo a WD kick, at least for stars that are not fully convective and that can evolve into red giants (M$_*>0.35$~M$_\odot$). The fraction of FFPs from WDs therefore grows with the age of the stellar population.
\end{enumerate}

\begin{figure}
    \centering
    \includegraphics[width=\linewidth]{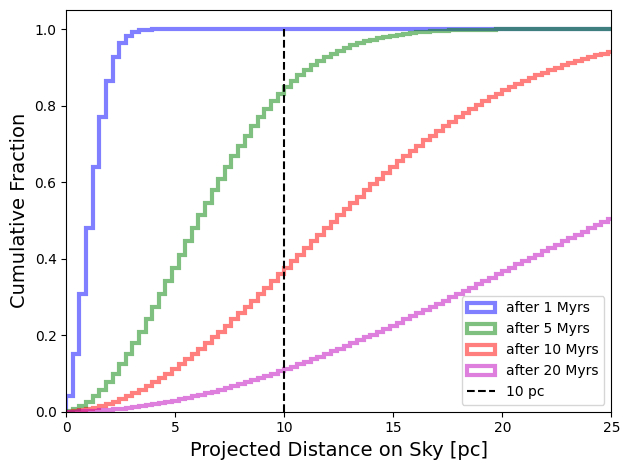}
    \caption{{\bf Time Evolution of WD-FFP Projected Distance Cumulative Distribution Function.} Shown is the cumulative likelihood to find a WD-kick ejected FFP within a given projected distance on the sky of its former host, for $1$, $5$, $10$, and $20$~Myr after \textcolor{black}{FFP ejection}. The dashed line highlights $10$~pc in projected distance, which can serve as a benchmark to limit the search space. See also Fig.~\ref{fig:5pc}.} 
    \label{fig:skytime}
\end{figure}

\begin{figure}
    \centering
    \includegraphics[width=\linewidth]{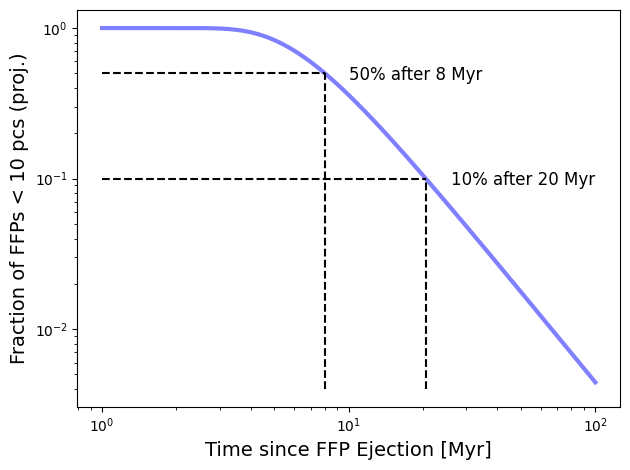}
    \caption{{\bf Time Evolution of Fraction of ejected FFPs within $10$~pc projected Sky Distance from the WD.} Shown is the likelihood to find a WD-kick ejected FFP within $10$~pc projected distance on the sky of its former host. After $8$~Myr, about $50\%$ of the ejected FFPs remain within this region, which drops to $10\%$ after about $20$~Myr.} 
    \label{fig:5pc}
\end{figure}

The combination of slow ejection speeds and enhanced planetary temperatures makes the regions surrounding young WDs very valuable targets for the search of WD-kick FFPs. The 3d-Maxwellian distribution for the ejection velocity shown in Eq.~\ref{eq:maxwell} becomes a 2d-Maxwellian when projected onto the surface of the sky. Given a most-likely excess ejection velocity of about $1.5$~pc/Myr as discussed in the previous sections, one can thus estimate a timeline for how long WD-ejected FFPs will remain in close projected proximity to their former host stars. Fig.~\ref{fig:skytime} shows the cumulative likelihood to find an ejected FFP within a certain projected sky distance from its former host, for various times following WD formation, \textcolor{black}{with the caveat that ejection might lag behind WD formation by a few Myr in the case of ejection via orbital instability}. When focusing on the closest $10$~pc in projected sky distance from a WD, to minimize contamination with the general FFP population, Fig.~\ref{fig:5pc} shows that half the ejected FFPs remain within this distance after about $8$~Myr, and still about a tenth after about $20$~Myr.

While these calculations show that a significant fraction of WD-kick FFPs will remain in close projected proximity to their hosts following ejection, given the prevalence of FFPs from other channels, one requires additional markers to identify this particular FFP population. Fortunately, the enhanced luminosity from a host star's AGB phase can dramatically alter the temperature and chemistry of its planets. As various previous studies have shown \citep[e.g.,][]{GinzburgSari2015,GinzburgSari2016,Lopez+2016,Jermyn+2017}, extended heating via enhanced stellar irradiation can lead to significant deep internal heating and radial inflation of gas giant planets. For example, Jupiter can be estimated to reach an equilibrium surface temperature of up to $700$~K during the sun's AGB phase (based on {\tt SSE} calculations of solar luminosity and mass loss evolution). The exact cooling curve of an exoplanet heated in this fashion after the enhanced stellar flux has ended depends strongly on the planet's internal structure and the duration of exposure to the increased stellar flux and a full exploration goes beyond the scope of this work. Nevertheless, here we attempt to make a simple estimate of the general temperature evolution based on considerations of re-heated and re-inflated gas giant planets. We note that this treatment is meant to be primarily qualitative.

\begin{figure}
    \centering
    \includegraphics[width=\linewidth]{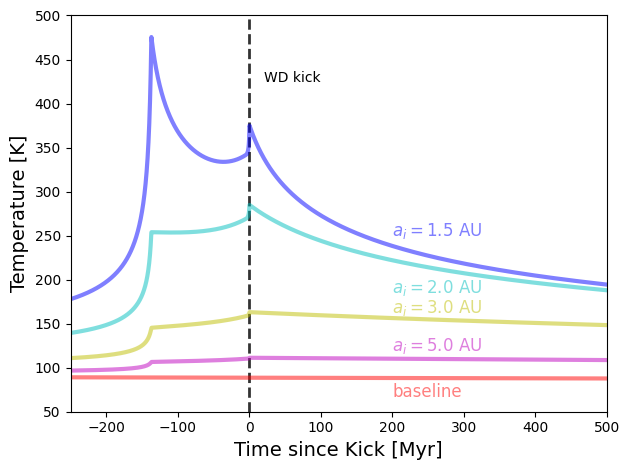}
    \caption{{\bf Temperature Evolution over Time of Jupiter-like FFPs heated by Red Giant Radiation before Ejection.} Shown is the approximate temperature evolution of Jupiter-like FFPs, assuming they initially orbited a Sun-like host star with initial SMAs of $1.5$~AU (blue curve), $2$~AU (cyan), $3$~AU (yellow), or $5$~AU (magenta). Ejection is assumed to have occurred at the end of the AGB phase, either through the WD kick directly or orbital instability. The unheated expected temperature of an approximately $10$~Gyr old Jupiter-like planet is shown in red as comparison. The planets retain an elevated temperature for many millions of years. The stellar mass and luminosity evolution was calculated using the stellar evolution code {\tt SSE} \citep{Hurley+00}. On the order of $2\%$ of the stellar energy flux was estimated to reach the planetary interior heat reservoir.}
    \label{fig:temp}
\end{figure}

The long-term intrinsic luminosity $L_P$ of a gas giant over time $t$ is, to first order, the result of the thermal cooling of the ions in its envelope once is has become mostly incompressible, as described by \citet{Guillot2005} with the equation \begin{equation}
    \label{eq:lum_cool}
    L_P=\eta\frac{G {M_P}^2}{R_P t},
\end{equation} with $M_P$ and $R_P$ being the planet's mass and radius, respectively, and $\eta$ a factor on the order of $0.01$ to $0.03$. Here, $\eta G {M_P}^2 /R_P$ essentially represents the initial heat reservoir $E_i$, which is a small fraction of the total gravitational binding energy of the planet. One can therefore also express the luminosity as a function of the remaining heat reservoir $E_H$, in the form of \begin{equation}
    \label{eq:lum_res}
    L_P=L_i\times \exp{\left(\frac{E_H}{E_i}-1\right)},
\end{equation} with some initial luminosity $L_i$. Thus, as the planet radiates away energy over time, the remaining internal heat reservoir $E_H$ is reduced, lowering the luminosity and temperature of the planet. However, the planet will also receive energy from outside sources; in particular, as its host star becomes a red giant, stellar radiation can increase tremendously, which has been studied in the literature in the context of re-inflated Jupiters \citep[e.g.,][]{GinzburgSari2015,GinzburgSari2016,Lopez+2016,Jermyn+2017}. Generally, while most of the incoming stellar radiation is only expected to heat the planetary surface layers up to the new equilibrium temperature, a small fraction of the incident energy (estimated to be on the order of a few percent; here we assume $\sim2\%$) is expected to reach into the deeper planetary regions, adding to the thermal reservoir $E_H$ of the overall planet. Depending on the initial orbit of the planet and the mass and luminosity evolution of the star, this additional heating can be quite significant and raise the intrinsic temperature by several hundred Kelvin, for many millions of years after the ejection of the planet as a FFP, as shown in Fig.~\ref{fig:temp}. As such, the regions surrounding young WDs can be expected to have an enhanced likelihood to be populated by heated FFPs, enhancing the detection likelihood through, for example, direct imaging infrared surveys.




\section{Summary \& Conclusions}

This work explored the potential for FFPs to be created by ejecting planets from evolved stellar systems via the recoil kick experienced by WDs during their formation from AGB stars. While for most known exoplanets the observed WD kick velocities are too small to lead to a significant number of direct planetary ejections, the orbital parameter changes caused by kicks can be expected to lead to orbital instabilities and eventual planetary ejections for nearly half of the known long-period multi-exoplanet systems. Additionally, very long period planets, which the currently dominant detection methods have difficulties in finding, are expected to exist for a significant fraction of stellar hosts, and have a much higher likelihood at being ejected by kicks, justifying the estimate that most currently existing WDs have ejected planets in their past. Overall, WD kicks are not the dominant channel for FFP production, as other pathways are expected to have much higher yields per stellar host. However, FFPs ejected via WD kicks are unique as they will have undergone extended evolution within a planetary system environment and will have undergone potentially significant heating from their host stars' red giant phase, making such FFPs potentially interesting targets for direct observation. FFPs from WD kicks will also only drift away from their origin systems at low speeds of, on average, $1.5$~pc/Myr, potentially allowing one to find warm FFPs at short projected distances from very young WDs. This unique FFP sub-population produced via WD kicks may be an important element for understanding the overall FFP demographics. While the upcoming {\it Roman} Galactic Exoplanet Survey is expected to detect a significant number of FFPs via microlensing, it cannot provide any information regarding the planetary temperatures or their ages and origins. As such, future observational campaigns should consider targeting the environment surrounding young WDs for direct detection of heated FFPs to complement the results of the microlensing survey.


\section*{Acknowledgments}
We thank the anonymous reviewer for their useful comments that have aided in improving this manuscript.  A.P.S.~acknowledges support from NASA grant 80NSSC24M0022. The authors are thankful for useful comments by and discussions with William DeRocco and Jessica Schonhut-Stasik.

\software{Matplotlib \citep{Hunter2007}, NumPy \citep{Harris+2020}, SciPy \citep{Virtanen+2020}, REBOUND \citep{ReinLiu2012,ReinSpiegel2015}, SSE \citep{Hurley+00}}

\label{References}

\bibliographystyle{aasjournal}
\bibliography{bibliography} 


\end{document}